\documentclass[twocolumn]{aastex701}
\usepackage{amsmath}
\usepackage{epsf}
\usepackage[dvipsnames]{xcolor}
\usepackage{xcolor}
\usepackage{enumitem}
\usepackage{mathtools}
\usepackage[mathcal]{eucal}
 \let\mathscr\relax
\usepackage[scr]{rsfso}
\usepackage{csvsimple}
\usepackage{booktabs}

\shorttitle{Hunting Wandering SMBHs using Emission Line Diagnostics}
\shortauthors{Thakurdesai et al.}

\def\arcs{\hbox{$^{\prime\prime}$}}

\begin{document}
\title{Hunting Wandering 3$<$z$<$8 Black Holes: 

Spatial Offsets in Ionization Ratio and Continuum Emission}

\author[0009-0008-9556-3821]{Urvi Thakurdesai}
\email{urvitd@hotmail.com}
\affiliation{Department of Astronomy, The University of Texas at Austin, Austin, TX, USA}

\author[0000-0003-1282-7454]{Anthony J. Taylor}
\email{anthony.taylor@austin.utexas.edu}
\affiliation{Department of Astronomy, The University of Texas at Austin, Austin, TX, USA}
\affiliation{Cosmic Frontier Center, The University of Texas at Austin, Austin, TX, USA}

\author[0000-0001-8519-1130]{Steven L. Finkelstein}
\email{stevenf@astro.as.utexas.edu}
\affiliation{Department of Astronomy, The University of Texas at Austin, Austin, TX, USA}
\affiliation{Cosmic Frontier Center, The University of Texas at Austin, Austin, TX, USA}

\author[0000-0002-9393-6507]{Gene C. K. Leung}
\email{gckleung@mit.ude}
\affiliation{MIT Kavli Institute for Astrophysics and Space Research, 77 Massachusetts Ave., Cambridge, MA 02139, USA}

\author[0000-0003-2332-5505]{\'Oscar A. Ch\'avez Ortiz}
\email{chavezoscar009@utexas.edu}
\affiliation{Department of Astronomy, The University of Texas at Austin, Austin, TX, USA}

\author[0000-0002-1410-0470]{Jonathan R. Trump}
\email{jonathan.trump@uconn.edu}
\affiliation{Department of Physics, 196 Auditorium Road, Unit 3046, University of Connecticut, Storrs, CT 06269, USA}

\author[0000-0001-8534-7502]{Bren E. Backhaus}
\email{b730b599@ku.edu}
\affiliation{Department of Physics and Astronomy, University of Kansas, Lawrence, KS, USA}

\author[0000-0001-7151-009X]{Nikko J. Cleri}
\affiliation{Department of Astronomy and Astrophysics, The Pennsylvania State University, University Park, PA 16802, USA}
\affiliation{Institute for Computational and Data Sciences, The Pennsylvania State University, University Park, PA 16802, USA}
\affiliation{Institute for Gravitation and the Cosmos, The Pennsylvania State University, University Park, PA 16802, USA}
\email{cleri@psu.edu}

\author[0000-0003-2388-8172]{Francesco D’Eugenio}
\email{francesco.deugenio@gmail.com}
\affiliation{Kavli Institute for Cosmology, University of Cambridge, Madingley Road, Cambridge, CB3 OHA, UK}

\author[0000-0001-9879-7780]{Fabio Pacucci}
\email{fabio.pacucci@cfa.harvard.edu}
\affiliation{Center for Astrophysics $\vert$ Harvard \& Smithsonian, Cambridge, MA 02138, USA}

\author[0000-0002-6610-2048]{Anton M. Koekemoer}
\email{koekemoer@stsci.edu}
\affiliation{Space Telescope Science Institute, 3700 San Martin Drive, Baltimore, MD 21218, USA}

\author[0000-0002-7959-8783]{Pablo Arrabal Haro}
\email{parrabalh@gmail.com}
\affiliation{Astrophysics Science Division, NASA Goddard Space Flight Center, 8800 Greenbelt Rd, Greenbelt, MD 20771, USA}

\author[0000-0002-9921-9218]{Micaela Bagley}
\email{mbagley@utexas.edu}
\affiliation{Astrophysics Science Division, NASA Goddard Space Flight Center, 8800 Greenbelt Rd, Greenbelt, MD 20771, USA}

\author[0000-0001-5414-5131]{Mark Dickinson}
\email{mark.dickinson@noirlab.edu}
\affiliation{NSF’s National Optical-Infrared Astronomy Research Laboratory, 950 North Cherry Avenue, Tucson, AZ 85719, USA}

\author[0000-0001-9187-3605]{Jeyhan Kartaltepe}
\email{jeyhan@astro.rit.edu}
\affiliation{Laboratory for Multiwavelength Astrophysics, School of Physics and Astronomy, Rochester Institute of Technology, 84 Lomb Memorial Drive, Rochester, NY 14623, USA}

\author[0000-0001-7503-8482]{Casey Papovich}
\email{papovich@tamu.edu}
\affiliation{Department of Physics and Astronomy, Texas A\&M University, College Station, TX, 77843-4242, USA}

\author[0000-0003-3382-5941]{Nor Pirzkal}
\email{npirzkal@stsci.edu}
\affiliation{ESA/AURA Space Telescope Science Institute, 3700 San Martin Drive, Baltimore, MD, 212129, USA}

\begin{abstract}
The early growth and assembly of supermassive black holes (SMBHs) remain key topics of interest in galaxy evolution. One of the scenarios predicted by theoretical models is that frequent minor mergers and asymmetric gas inflows may cause SMBHs to temporarily reside off-center within their host galaxies in the early universe. To observationally test this scenario, we investigate whether spatially offset ionization signatures---which may be indicative of active galactic nuclei (AGN)---can be identified. Using JWST NIRSpec PRISM spectroscopy from the Cosmic Evolution Early Release Science (CEERS) survey, we analyze the 2D spectra of 90 high-redshift galaxies $(3<z<8)$, including two known broad-line AGN. By measuring key emission lines such as H$\alpha$, {\hbox{{\rm H}\kern 0.05em$\beta$}}, {\hbox{{\rm [O}\kern 0.1em{\sc iii}{\rm ]}}}$\lambda$5007, {\hbox{{\rm [Ne}\kern 0.1em{\sc iii}{\rm ]}}}${\lambda 3868}$, and {\hbox{{\rm [O}\kern 0.1em{\sc ii}{\rm ]}}}${\lambda\lambda 3727,3729}$ we derive spatial flux ratio profiles, and focus on {\hbox{{\rm [O}\kern 0.1em{\sc iii}{\rm ]}}}/{\hbox{{\rm H}\kern 0.05em$\beta$}} as a tracer of high-ionization mechanisms that may indicate AGN activity.
We identify 26 galaxies ($\sim30\%$ of the sample) with significant localized peaks in {\hbox{{\rm [O}\kern 0.1em{\sc iii}{\rm ]}}}/{\hbox{{\rm H}\kern 0.05em$\beta$}}. Out of these 26 galaxies, 12 sources ($\sim46\%$) exhibit significant spatial offsets between the peak {\hbox{{\rm [O}\kern 0.1em{\sc iii}{\rm ]}}}/{\hbox{{\rm H}\kern 0.05em$\beta$}} ratio and the stellar continuum center. Six of these sources show the highest amount ($>1.5$) pixel spatial offsets. 
This spatial offset between ionization structure and stellar centers offers a promising avenue to probe early SMBH evolution and its connection to galaxy formation.

\end{abstract}

\section{Introduction}

Supermassive black holes (SMBHs) have long been a subject of interest in astrophysics due to their significant role in regulating the growth and evolution of galaxies \citep[e.g.,][]{BH_growth_simulation,BH_growth_simulation_2}. While correlations such as the relation between black-hole mass ($M_{\rm BH}$) and velocity dispersion ($\sigma_*$) suggest a co-evolution between SMBHs and their host galaxies (e.g., \citealp{Ferrarese2000, Gebhardt2000}) with indications that SMBHs can regulate star formation and that stellar feedback can influence black hole growth, the epoch at which these correlations were first established---particularly in the early universe---remains uncertain. A key unresolved question is how SMBHs grow during the formative stages of galaxy assembly, especially at redshifts $z > 3$, when frequent mergers and rapid gas accretion dominate galaxy dynamics \citep{mergers_SMBH, merger_bh}. In particular, the concept of galaxy–black hole co-evolution has emerged as a foundational framework for understanding the interdependent formation and development of central black holes and their host galaxies, especially in the context of the early universe where such processes are less well-constrained \citep[e.g.,][]{SMBH_contraints_1, SMBH_constraints_2, SMBH_constraints_6, SMBH_constraints_4,
BH_growth_simulation,  SMBH_constraints_7, SMBH_constraints_5, SMBH_constraints_8, SMBH_constraints_9, SMBH_constraints_10, trump, SMBH_constraints_3,  Backhaus, SMBH_constraints_11, SMBH_constraints_12}.

Simulations of early SMBH growth predict that black holes may occasionally reside off-center within their host galaxies during both the early and late stages of cosmic time as a consequence of hierarchical galaxy assembly and long dynamical sinking timescales \citep[e.g.,][]{tremmel_2018_wander, tremmel_2018MNRAS.475.4967T, priya_smbh_wander, Bellovary_2021, weller_bh_wandering, agn_wander_jeon}. Dynamical processes such as minor mergers, uneven gas inflows, long dynamical friction timescales, and gravitational recoil kicks following black hole mergers can displace SMBHs from the galactic nucleus, resulting in a wandering" phase that may persist for a significant period of time \citep{Volonteri2010, VolonteriMadau2008, tremmel_2018_wander, Bechmann_2023, DiMatteo_2023_MBHs, DiMatteo_2023_imbh}. These off-nuclear SMBHs may continue actively accreting until they run out of fuel, producing AGN signatures that ionize their local environment \citep{greene_offset_AGN_accreting, chen_2023, chen_2024, Saeedzadeh_2024}. However, observationally detecting such spatial offsets is challenging---particularly at high redshift---due to the relatively small sizes of galaxies coupled with limitations in instrument angular resolution and sensitivity.

Recent simulation studies have attempted to quantify both the mass range and typical displacement of wandering black holes, suggesting that they are often less massive than $M_{\rm BH} \lesssim 10^6 -10^7M_\odot$, and can reach offsets of several hundred parsecs within their host galaxies \citep[e.g.,][]{ricarte2021_imbh_mergers, agn_wander_jeon, sinking_smbh}. These results motivate the observational search for off-nuclear AGNs as a potential probe of early SMBH growth and dynamical evolution.

The advent of JWST provides a powerful opportunity to address this challenge. In particular, the Near Infrared Spectrograph (NIRSpec) PRISM mode enables spatially resolved spectroscopy of key diagnostic emission lines, including [OIII]$\lambda5007$ (hereafter [OIII]), {\hbox{{\rm H}\kern 0.05em$\beta$}}, H$\alpha$, {\hbox{{\rm [Ne}\kern 0.1em{\sc iii}{\rm ]}}}${\lambda 3868}$, and {\hbox{{\rm [O}\kern 0.1em{\sc ii}{\rm ]}}}${\lambda\lambda 3727,3729}$, for galaxies at high redshift ($1<z<10$) \citep{nirspec}. Among these, the {\hbox{{\rm [O}\kern 0.1em{\sc iii}{\rm ]}}}/{\hbox{{\rm H}\kern 0.05em$\beta$}} flux ratio, especially higher levels, is a well-established tracer of AGN-driven ionization especially at low-$z$ for higher metallicity systems  \citep{bpt81, trump}, allowing spatial maps of ionized regions to be directly compared with the stellar continuum morphology. 

Measurements of this ratio, particularly when centered on or near the galaxy nucleus \citep{Kewley_lineratios, kauffman_lineratios, trump_lineratios, kewley_lineratios_review}, provide strong constraints on the strength and spatial extent of AGN-driven ionization at $z \lesssim 2$, offering insight into the role of AGNs in regulating star formation and shaping host galaxy evolution. At higher redshift, however, the situation is more complex. Many star-forming galaxies occupy the nominal AGN regions of classical diagnostic diagrams (e.g., BPT, OHNO), largely due to the harder radiation fields and lower metallicities that characterize high-$z$ star formation \citep[e.g.,][]{bren_ratios_1,   nikko_ionization, sanders_ion_excite, trump_ratios_3,bren_ratios_2}. As a result, AGN and star-forming galaxies are often indistinguishable in these line-ratio spaces. Moreover, spatially offset ionized regions may also arise from clumpy star formation within gas-rich high-redshift galaxies, potentially tracing giant star-forming clumps formed through disk instabilities \citep{dekel_2009, guo_2012}. Distinguishing between AGN-driven ionization and clump-dominated star formation therefore remains a key challenge when interpreting spatially resolved emission-line structure in the early Universe.

JWST spectroscopy offers unprecedented spatial resolution, enabling detailed studies of the internal structure of galaxies in the early Universe. The NIRSpec IFU mode is optimal for exploring the two-dimensional projected spatial structure of galaxies, and recent JWST/NIRSpec IFU observations have revealed spatially offset broad-line AGN components at $z>7$, interpreted as evidence for dual or merging black holes in the early Universe \citep{AGN_offset_GA_NIFS}. These results demonstrate the power of spatially resolved spectroscopy in uncovering complex black hole and galaxy evolution at high redshift. While the IFU is ideal for spatially resolved spectroscopy, the observations are limited to small, targeted samples. In contrast, the extensive archival NIRSpec Multi-Object Spectroscopy data provide spectra for large numbers of galaxies and retain spatial information in the cross-dispersion direction, which contain untapped potential for spatial studies. 

In this work, we investigate whether localized elevated ionization regions, traced by spatially resolved {\hbox{{\rm [O}\kern 0.1em{\sc iii}{\rm ]}}}/{\hbox{{\rm H}\kern 0.05em$\beta$}} profiles, are spatially offset from the stellar light-weighted centers of high-redshift galaxies. Using data from the CEERS survey \citep{CEERS2025}, we analyze the 2D NIRSpec PRISM spectra for 90 galaxies, including known broad-line AGNs \citep[BLAGNs;][]{anthony}, to search for {\hbox{{\rm [O}\kern 0.1em{\sc iii}{\rm ]}}}/{\hbox{{\rm H}\kern 0.05em$\beta$}} peaks that do not coincide with the galaxy continuum center. Such spatial offsets could indicate the presence of off-nuclear AGNs, supporting the idea that early SMBH growth may be decoupled from central stellar mass assembly \citep{greene_offset_AGN_accreting, Tremmel2018, priya_smbh_wander} or---alternatively---arise from non-uniform star formation or asymmetric feedback. 

We organize the paper as follows. In Section \ref{sec:datasets}, we describe the dataset and outline how it was derived from the CEERS catalog. Section \ref{sec:techniques} details our 
procedures for line fitting and constructing spatial and flux ratio profiles. We also focus on identifying significant localized {\hbox{{\rm [O}\kern 0.1em{\sc iii}{\rm ]}}}/{\hbox{{\rm H}\kern 0.05em$\beta$}} peaks. In Section \ref{sec:results}, we calculate the spatial offsets between the {\hbox{{\rm [O}\kern 0.1em{\sc iii}{\rm ]}}}/{\hbox{{\rm H}\kern 0.05em$\beta$}} peak and the galaxy continuum center and present our findings, focusing on  correlations between flux ratios and their significant localized {\hbox{{\rm [O}\kern 0.1em{\sc iii}{\rm ]}}}/{\hbox{{\rm H}\kern 0.05em$\beta$}} peaks. We further explore potential connections between these localized peaks and AGN activity using the OHNO diagnostic plot. Finally, Section \ref{sec:summary} summarizes our results and conclusions.

We assume standard cosmology conditions throughout the paper: $\Omega_{m} = 0.3$, $\Omega_{\Lambda} = 0.7$, and $H_0 = 70 \ \rm km \ \rm s^{-1} \ Mpc^{-1}$. 

\section{Data Sets} \label{sec:datasets}

Our initial sample comprises 1089 high-redshift galaxies in the redshift range $1 < z < 9$, observed with the Near Infrared Spectrograph instrument onboard the James Webb Space Telescope as part of Cycle 1 Early Release Science (ERS) program, Cosmic Evolution Early Release Science Survey (CEERS $\#1345$; \citealt{CEERS2025}). These sources are observed with the Multi-Object Spectroscopy mode, using the PRISM/CLEAR filter configuration, which provides low-resolution ($\rm R \sim 30-330$) slit-based spectroscopy with wavelength coverage of $0.6 -5.3 \ \mu m$ and a cross-dispersion plate scale of $0.1\arcs$/pixel \citep{nirspec}. The data spans over six pointings across the CEERS Extended Groth Strip (EGS) field. 

We use a collaboration-wide manually vetted catalog of redshifts for these CEERS sources and we select only those with the highest quality grade \citep{pablo_reduc}. We further impose a minimum redshift of $z=3$, ensuring that the emission lines of interest are sufficiently resolved in the PRISM observations. This results in a sample of 208 spectroscopically confirmed galaxies.

\begin{figure}[t]
    \centering
    \includegraphics[width=0.6\columnwidth]{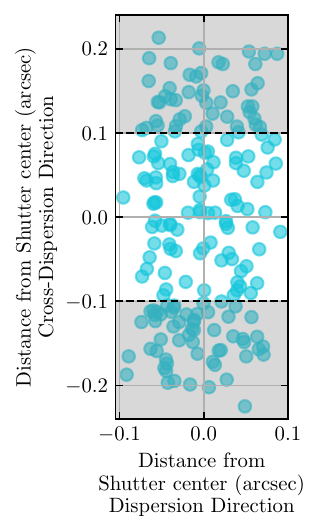}
    \caption{
        Intra-shutter position of 195 sources from the visually inspected CEERS sample in the redshift range $3 < z < 8$, shown in corrected slit-based coordinates using JWST slit positions (in arcseconds). We apply a positional cut of $\sim 0.1\arcsec$ along the y-axis to retain sources centrally located within the slit. Grey regions indicate sources excluded by this criterion. This cut results in a sample of 90 galaxies and ensures that only sources with minimal effects from slit edges and nodded backgrounds are included in the final analysis, while also maximizing the usable off-center region of each galaxy for our further analysis.
    }
\label{fig:central_slit_cut}
\end{figure}

Each spectrum in the original sample was reduced using a custom non-nodded background subtraction, ensuring data quality suitable for spatially resolved analysis \citep{pablo_reduc}. Unlike the standard procedure of combining and subtracting nodded pairs along the slit, we adopted non-nodded background subtraction to avoid truncation of the 2D spectra that can occur when nod distances are comparable to the galaxy sizes. This approach also removes residual negatives from the nodded subtraction that would otherwise distort the spatial profile. We begin our selection by visually inspecting the 2D spectra of the 208 galaxies in this sample to exclude poor-quality data affected by contamination or truncation. This step removes sources for which accurate spatial analysis would be compromised due to artifacts such as hot pixels or companion sources and gives us a sample of 195 sources.

To ensure that we extract the maximum amount of spatial information around the centroid of each galaxy, and that the sources are minimally affected by slit loss and the bar shadow, we apply a slit-position-based cut on the spatial location of each galaxy within the aperture.
We restrict our analysis to galaxies located within $0.1\arcs$ (one NIRSpec pixel) of the center of the slit in the cross-dispersion direction as shown in Figure \ref{fig:central_slit_cut} \citep{nirspec}. This positional filtering minimizes the influence of slit edge effects and maximizes the signal-to-noise ratio (S/N) of the spatially resolved line ratio measurements.

\begin{figure}[t]
    \includegraphics[width=\columnwidth]{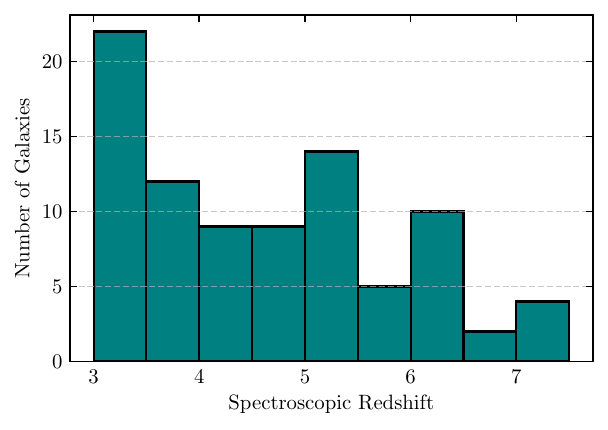}
    \caption{Distribution of spectroscopic redshifts for our reduced sample of 90 galaxies that are centrally positioned within the JWST/NIRSpec slit, selected using a positional cut of approximately $0.1\arcs$ along the spatial axis of the slit. This cut ensures optimal placement within the slit for reliable spatially resolved spectroscopy.  The redshifts are binned by $\Delta z = 0.5$ across the range $3 < z < 8$.}

\label{fig:redshift_dist}
\end{figure}

Applying this selection criterion results in a final sample of 90 galaxies. Figure~\ref{fig:central_slit_cut} presents an overview of our slit-position cut and the sample curation process, while Figure~\ref{fig:redshift_dist} shows the redshift distribution of the final selected targets.

A known artifact of the NIRSpec MSA observations when non-nodded background subtraction is used is the appearance of regions of negative flux, which result from the bar-shadow---the occluding of the detector by the bars that separate the micro-shutters in the MSA. 
\citep{nirspec}. While the JWST Calibration Pipeline offers a bar-shadow correction suitable for extended sources, this correction is far from perfect and is prone to producing artifacts in processed data. We therefore neglect the bar-shadow correction and instead extract the central five spatial pixels of each source that are least affected by these artifacts. These are identified by locating the central pixel of an object's continuum emission and including two adjacent pixels on either side. This approach ensures that our analysis focuses on the regions of highest S/N and minimizing the effects of the bar-shadow. 

To mitigate the impact of the bar-shadow on our spatial profile measurements, we assess the compactness of the galaxies in our sample using their half-light radii in the F200W filter (F200W $R_h$). We do this with photometric catalogs from the UNICORN project (S. Finkelstein et al. in prep), which include all {\it HST}/ACS and {\it JWST}/NIRCam data in this fields. The catalog is broadly based on the photometric procedures of \citep{finkelstein24}, thus we refer the reader there for more information. We find that 51 out of those 90 galaxies in the sample ($>50\%$) which are covered by NIRCam coverage have F200W $R_h$ $\leq 0.2\arcs$, with a majority $< 0.1\arcs$. These sizes are small compared to the slit width and bar-shadow scale, hence satisfying the compactness criterion, indicating that their spatial profiles are dominated by compact emission and are therefore only minimally affected by partial obscuration. We do not exclude more extended sources; rather, this analysis demonstrates that the majority of the sample is sufficiently compact that bar-shadow effects are unlikely to significantly bias our spatial measurements. Given that more than half of the sample satisfies this compactness criterion, we assume it is representative of the full set of 90 sources.

To also identify potential AGN hosts within our final sample, we cross-match the galaxy coordinates with the BLAGN census provided in \citet{anthony}, which compiled broad-line AGN identifications across multiple deep extragalactic fields, including all of CEERS. This enables us to flag two known AGN sources (CEERS MSA ID: 746, 2782) in our dataset for further comparison with spatial ionization diagnostics. 

For each galaxy, we analyze both 1D spectra and 2D spatially resolved spectra from the abovementioned NIRSpec observations, allowing for detailed analysis of emission line diagnostics. These data products form the basis for the ionization ratio mapping and AGN offset studies presented in subsequent sections.

\begin{figure*}
    \includegraphics[width=\textwidth]{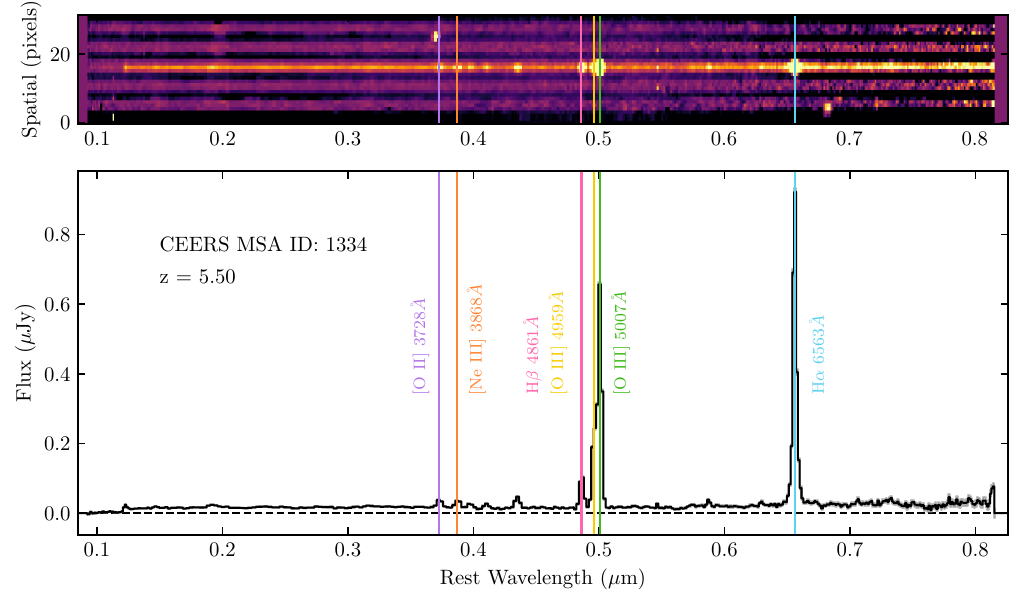}

    \caption{\textbf{Top:} The colormap shows the 2D spectrum of a source from our sample, CEERS MSA ID: 1334. The y-axis represents the spatial dispersion of the galaxy. The x-axis represents the spectral dispersion of the galaxy (in units of $\mu$m; converted to rest-frame using the known spectroscopic redshift) which highlights emission features of the galaxy. \textbf{Bottom:} The black line represents the extracted 1D spectrum of the same source with $1\sigma$ uncertainty shaded in gray. The key emission lines of {\hbox{{\rm [O}\kern 0.1em{\sc iii}{\rm ]}}}$\lambda5007$, {\hbox{{\rm [O}\kern 0.1em{\sc iii}{\rm ]}}}$\lambda4959$, H$\beta\lambda4861$, H$\alpha\lambda6563$, {\hbox{{\rm [Ne}\kern 0.1em{\sc iii}{\rm ]}}}$\lambda3868$, {\hbox{{\rm [O}\kern 0.1em{\sc ii}{\rm ]}}}$\lambda3728$ are also indicated by different colored lines. We extract 1D spectra from the five central spatial rows and measure the emission-line fluxes using emission-line fitting.}
\label{fig:1D_2D_spectra}
\end{figure*}

\section{Analysis Techniques} \label{sec:techniques}

In this section, we describe the methodology used to construct spatially resolved flux ratio profiles from our sample of 90 high-redshift galaxies. The objective is to identify ionization signatures associated with AGN activity and to examine their spatial distribution relative to the galactic center.

\subsection{Spatial Line Fitting}

We begin by identifying prominent emission features in the NIRSpec data, with a particular focus on emission lines that can be used to trace AGN activity. Specifically, we target H$\alpha$, {\hbox{{\rm H}\kern 0.05em$\beta$}}, {\hbox{{\rm [O}\kern 0.1em{\sc iii}{\rm ]}}}${\lambda 5007}$, {\hbox{{\rm [O}\kern 0.1em{\sc iii}{\rm ]}}}${\lambda 4959}$, {\hbox{{\rm [Ne}\kern 0.1em{\sc iii}{\rm ]}}}${\lambda 3868}$, and {\hbox{{\rm [O}\kern 0.1em{\sc ii}{\rm ]}}}${\lambda 3278}$  as our key emission lines (the {\hbox{{\rm [O}\kern 0.1em{\sc ii}{\rm ]}}} doublet $\lambda$$\lambda$3726, 3729 is unresolved in the PRISM observations; therefore, we adopt a single emission line at $3728$\AA{} as a representative {\hbox{{\rm [O}\kern 0.1em{\sc ii}{\rm ]}}} emission feature to fit with a single Gaussian, see below). 

These lines are well-established tracers of both star formation and AGN ionization. At low redshift, their ratios—particularly [O III]/{\hbox{{\rm H}\kern 0.05em$\beta$}}—serve as powerful diagnostics of different ionizing sources \citep{trump, Backhaus_OHNO, Backhaus}. At higher redshifts ($z \gtrsim 2$), however, the distinction becomes more ambiguous, as typical star-forming galaxies often occupy the usual AGN regions of diagnostic diagrams due to harder radiation fields, higher densities, and lower metallicities \citep{ cameron_2023_ISM_conditions, bren_ratios_1, bren_ratios_2, sanders_ion_excite, trump_ratios_3, boyett_ion_excite, topping_2024_ion_radiation_fields, nikko_OHNO}. Nevertheless, measuring [O III]/{\hbox{{\rm H}\kern 0.05em$\beta$}} remains valuable for identifying possible offsets from the star-forming locus that could signal AGN activity, particularly in cases where additional signatures (e.g., broad emission lines or off-nuclear ionization) can provide corroborating evidence. 

Once the emission lines are located in the 1D spectra, we use their corresponding wavelengths to identify the spatial extent of these features in the 2D dispersed spectra. An example 1D and 2D spectrum of a single galaxy with marked key emission lines is shown in Figure \ref{fig:1D_2D_spectra}.

We perform line fitting to extract integrated fluxes for {\hbox{{\rm [O}\kern 0.1em{\sc iii}{\rm ]}}} and {\hbox{{\rm H}\kern 0.05em$\beta$}} from the 2D spectra, which are used in constructing spatially resolved ionization profiles. We also fit the {\hbox{{\rm [Ne}\kern 0.1em{\sc iii}{\rm ]}}}$\lambda3868$ and {\hbox{{\rm [O}\kern 0.1em{\sc ii}{\rm ]}}}$\lambda3728$ (hereafter, {\hbox{{\rm [Ne}\kern 0.1em{\sc iii}{\rm ]}}} and {\hbox{{\rm [O}\kern 0.1em{\sc ii}{\rm ]}}}, respectively) lines which we later use for emission line diagnostics. We extract 1D spectra from the five central spatial rows of the 2D spectrum and perform emission line fits on each individual 1D spectrum to measure the line fluxes of the emission lines.

\begin{figure}
    \includegraphics[width=\columnwidth]{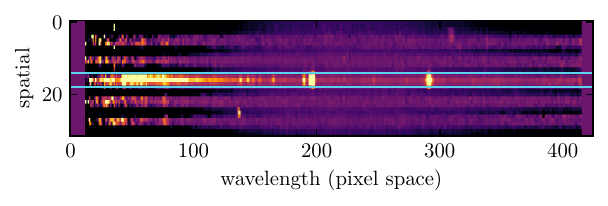}
    \includegraphics[width=\columnwidth]{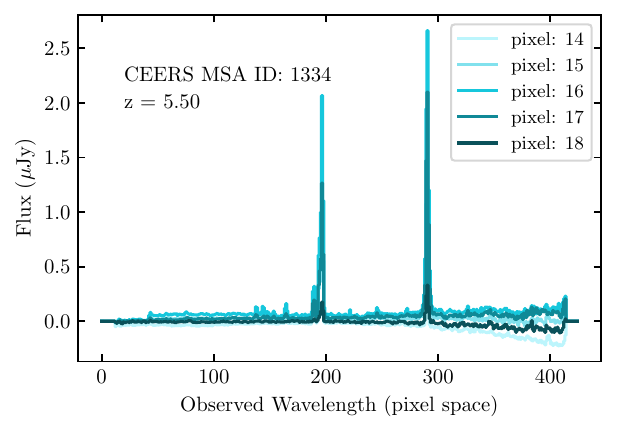}
    \includegraphics[width=\columnwidth]{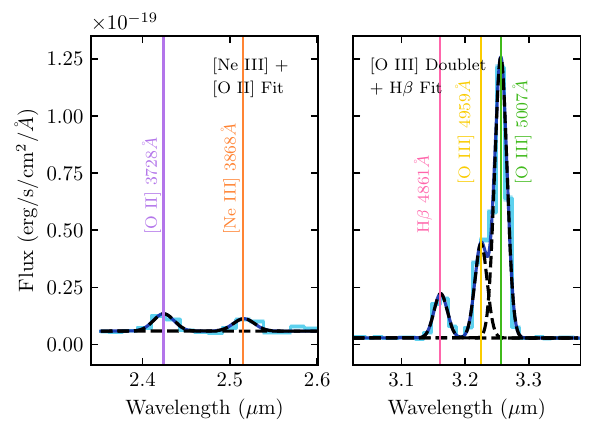}
    \caption{\textbf{Top:} An example 2D spectrum of source MSA ID: 1334. The blue lines represent the row by row region that we perform MCMC fits on to include central regions of the galaxy. \textbf{Middle:}  1D spectra for each of the five central row by row pixels. \textbf{Bottom:} An example of an MCMC fit applied to emission lines of the flux in the 1D spectrum.  We model the {\hbox{{\rm [O}\kern 0.1em{\sc iii}{\rm ]}}}$\lambda5007$, {\hbox{{\rm [O}\kern 0.1em{\sc iii}{\rm ]}}}$\lambda4959$, and H$\beta\lambda4861$ lines with a triple Gaussian profile, while {\hbox{{\rm [Ne}\kern 0.1em{\sc iii}{\rm ]}}}$\lambda3868$ and {\hbox{{\rm [O}\kern 0.1em{\sc ii}{\rm ]}}}$\lambda3728$ are each fit with a single Gaussian. The histogram represents the data and the black dashed lines represent the fit performed on each line. The dark blue line represents the combined fit from the black dashed lines. The MCMC sampling yields posterior distributions for the line parameters, allowing us to derive robust integrated fluxes used in our subsequent analysis.}
\label{fig:emceefit}
\end{figure}

We implement the fitting using a Markov Chain Monte Carlo (MCMC) framework from the \texttt{emcee} package, which allows for robust exploration of the parameter space and uncertainty estimation \citep{emcee2013}. We simultaneously fit three Gaussians for the {\hbox{{\rm [O}\kern 0.1em{\sc iii}{\rm ]}}} doublet and {\hbox{{\rm H}\kern 0.05em$\beta$}}, and two additional Gaussians for {\hbox{{\rm [Ne}\kern 0.1em{\sc iii}{\rm ]}}} and {\hbox{{\rm [O}\kern 0.1em{\sc ii}{\rm ]}}} using 11 free parameters: redshift, continuum flux density near {\hbox{{\rm [O}\kern 0.1em{\sc iii}{\rm ]}}}, {\hbox{{\rm [O}\kern 0.1em{\sc iii}{\rm ]}}} line width, {\hbox{{\rm H}\kern 0.05em$\beta$}} line width, {\hbox{{\rm [O}\kern 0.1em{\sc iii}{\rm ]}}}$\lambda$5007 line flux (the {\hbox{{\rm [O}\kern 0.1em{\sc iii}{\rm ]}}}$\lambda$4959 line flux is fixed to the {\hbox{{\rm [O}\kern 0.1em{\sc iii}{\rm ]}}}$\lambda$5007 line flux in a 1:2.98 ratio) \citep{OIII_ratio}, {\hbox{{\rm H}\kern 0.05em$\beta$}} line flux, continuum flux density near {\hbox{{\rm [O}\kern 0.1em{\sc ii}{\rm ]}}}, {\hbox{{\rm [O}\kern 0.1em{\sc ii}{\rm ]}}} line width, {\hbox{{\rm [Ne}\kern 0.1em{\sc iii}{\rm ]}}} line width, {\hbox{{\rm [O}\kern 0.1em{\sc ii}{\rm ]}}} line flux, and {\hbox{{\rm [Ne}\kern 0.1em{\sc iii}{\rm ]}}} line flux. 
We fit this model to each 1D spatial row spectrum with MCMC, using 32 walkers and 10000 steps, where the first $2/3$ of the steps are discarded to ensure that the walkers have converged on an optimal solution. The best fit parameters are taken as the medians of the posterior distributions from the MCMC fitting routine. An example MCMC fit is shown in Figure \ref{fig:emceefit}.

To ensure a reliable S/N before constructing flux ratio profiles, we apply a cut requiring $\rm SNR \geq 2$ for both the {\hbox{{\rm [O}\kern 0.1em{\sc iii}{\rm ]}}} and {\hbox{{\rm H}\kern 0.05em$\beta$}} emission lines. We find the SNR by using the median and 68\% confidence limit from the emcee posteriors. This criterion reduces our sample from 90 to 87 sources.

\subsection{Spatial Profiles and Flux Ratio Profiles}

\begin{figure*}
    \includegraphics[width=0.49\linewidth]{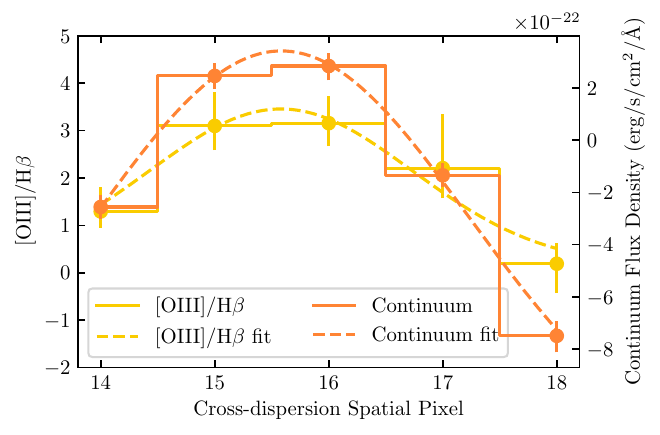}
    \includegraphics[width=0.49\linewidth]{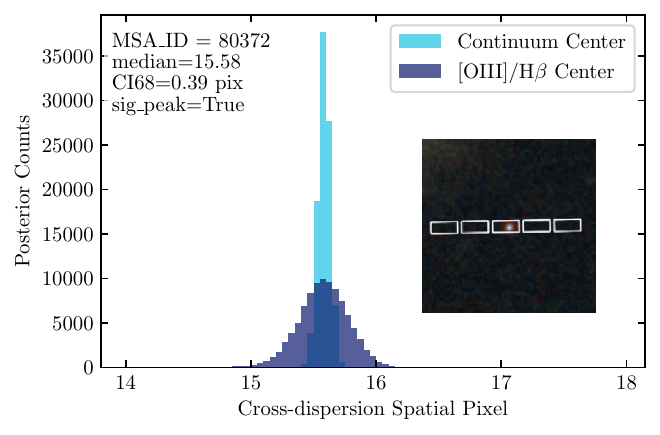}\\
    
    \includegraphics[width=0.49\linewidth]{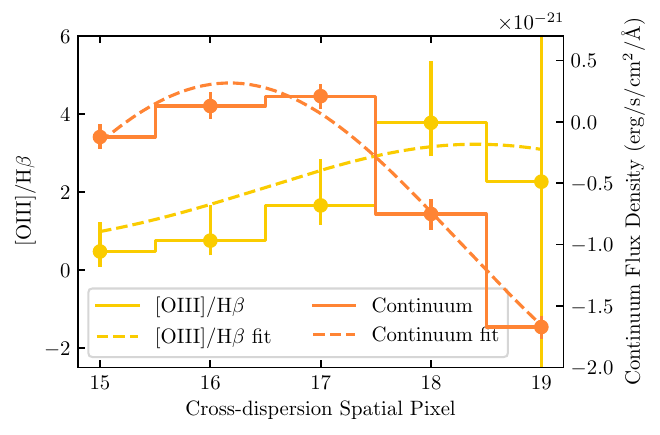}
    \includegraphics[width=0.49\linewidth]{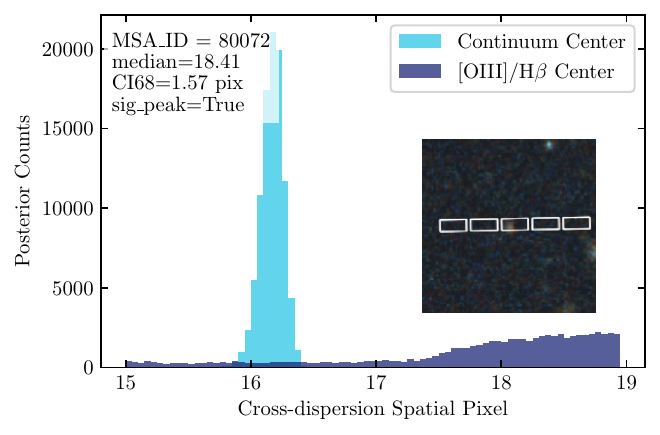}\\
    \includegraphics[width=0.49\linewidth]{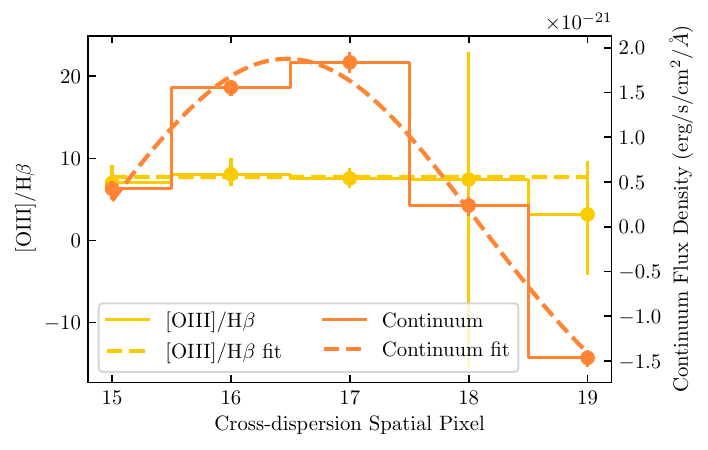}
    \includegraphics[width=0.49\linewidth]{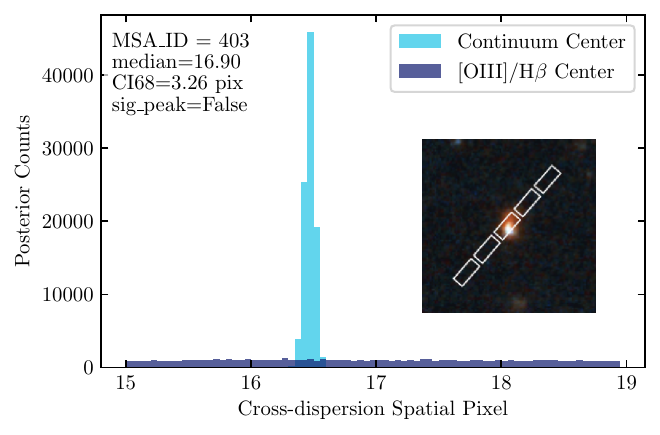}

    \caption{Examples of flux ratio spatial profiles, with fits (on the left) and their posterior distributions with NIRCam images (on the right) for three sources in our sample. \textbf{Left column:} The yellow distributions show the spatial variation of the {\hbox{{\rm [O}\kern 0.1 em{\sc iii}{\rm ]}}}/{\hbox{{\rm H}\kern 0.05 em$\beta$}} ionization ratio derived from MCMC fitting, while the orange distributions represent the corresponding continuum flux. The dashed lines for each color indicate the respective Gaussian fits which we apply further to the {\hbox{{\rm [O}\kern 0.1 em{\sc iii}{\rm ]}}}/{\hbox{{\rm H}\kern 0.05 em$\beta$}} ratio and the underlying continuum. \textbf{Right column:} The light blue distribution shows the posterior for the continuum central pixel, and the darker blue distribution shows the posterior for the {\hbox{{\rm [O}\kern 0.1em{\sc iii}{\rm ]}}}/{\hbox{{\rm H}\kern 0.05em$\beta$}} ionization ratio. The \texttt{median} represents the median pixel value of the flux-ratio posterior distribution. The \texttt{CI68} corresponds to the highest posterior density interval (HPDI) width enclosing the tightest $68\%$ of the {\hbox{{\rm [O}\kern 0.1em{\sc iii}{\rm ]}}}/{\hbox{{\rm H}\kern 0.05em$\beta$}} posterior distribution. The parameter \texttt{sig\_peak} indicates whether the flux-ratio peak is statistically significant relative to the HPDI width. \textbf{Top panel:} MSA ID 80372 shows well localized continuum and {\hbox{{\rm [O}\kern 0.1em{\sc iii}{\rm ]}}}/{\hbox{{\rm H}\kern 0.05em$\beta$}} posterior signifying statistically significant localized peaks that are spatially coincident. \textbf{Middle panel:} MSA ID 80072 also displays a significant localized enhancement in the {\hbox{{\rm [O}\kern 0.1em{\sc iii}{\rm ]}}}/{\hbox{{\rm H}\kern 0.05em$\beta$}} ratio. However, the separation between the posterior peaks of the continuum and the {\hbox{{\rm [O}\kern 0.1em{\sc iii}{\rm ]}}}/{\hbox{{\rm H}\kern 0.05em$\beta$}} ratio suggests that the region of strongest ionization is spatially offset from the galaxy’s center. \textbf{Bottom panel:} MSA ID 403 exhibits a relatively uniform {\hbox{{\rm [O}\kern 0.1em{\sc iii}{\rm ]}}}/{\hbox{{\rm H}\kern 0.05em$\beta$}} distribution, as seen in both the spatial and posterior profiles, indicating an overall uniform ionization distribution across the galaxy.
}
\label{fig:flux_ratio_profiles}
\end{figure*}

We next compute flux ratio profiles by dividing the integrated {\hbox{{\rm [O}\kern 0.1em{\sc iii}{\rm ]}}} flux obtained from the MCMC fits by the {\hbox{{\rm H}\kern 0.05em$\beta$}} flux at each spatial pixel as shown in Figure \ref{fig:flux_ratio_profiles}. This {\hbox{{\rm [O}\kern 0.1em{\sc iii}{\rm ]}}}/{\hbox{{\rm H}\kern 0.05em$\beta$}} ratio, which we refer to as the ionization ratio hereafter, is a key diagnostic for identifying AGN-driven ionization. Significantly elevated values of this ratio in a spatially localized region often indicate stronger ionizing radiation regions, which could possibly be characteristic of AGN activity \citep{trump}. We do not perform dust corrections as these two emission lines lie close enough to each other that the corrections are negligible. 

To identify the spatial center of the galaxy’s stellar component, as part of of MCMC fitting above, we fit the continuum profile over line-free regions in rest-frame $4650$–$5100$~\AA~for {\hbox{{\rm [O}\kern 0.1em{\sc iii}{\rm ]}}}$\lambda5007$ and {\hbox{{\rm H}\kern 0.05em$\beta$}}, and $3600$–$4000$~\AA~for {\hbox{{\rm [Ne}\kern 0.1em{\sc iii}{\rm ]}}} and {\hbox{{\rm [O}\kern 0.1em{\sc ii}{\rm ]}}}. This fitted continuum profile, which traces the stellar component, is overplotted on the flux ratio profiles in Figure \ref{fig:flux_ratio_profiles}. The peak of this continuum emission serves as a proxy for the galaxy’s stellar morphological center and is also defined as the center of our five central pixel region. By comparing the position of the ionization peak in the {\hbox{{\rm [O}\kern 0.1em{\sc iii}{\rm ]}}}/{\hbox{{\rm H}\kern 0.05em$\beta$}} profile with the continuum peak, we measure spatial offsets that may be indicative of potential AGN wandering or asymmetric feedback.

\subsection{Significant Ionization Peaks in {\hbox{{\rm [O}\kern 0.1em{\sc iii}{\rm ]}}}/{\hbox{{\rm H}\kern 0.05em$\beta$}}:} 

Within our reduced sample of 87 galaxies, we aim to identify significant enhancements in the {\hbox{{\rm [O}\kern 0.1em{\sc iii}{\rm ]}}}/{\hbox{{\rm H}\kern 0.05em$\beta$}} flux ratio. These enhancements typically appear as localized peaks in the flux ratio profiles—sometimes centered within the galaxy, but often offset from the nucleus—and show a tapering behavior toward the outer regions, consistent with expectations for either AGN-driven ionization \citep[e.g.,][]{trump} or compact off-nuclear star-forming regions, as illustrated in the top and middle panel of Figure \ref{fig:flux_ratio_profiles}. The observed concentration and smoothly declining wings of these peaks suggest that the underlying ionizing source is both intense and spatially compact, indicative of AGN activity or concentrated star formation.

To statistically quantify the significance of these {\hbox{{\rm [O}\kern 0.1em{\sc iii}{\rm ]}}}/{\hbox{{\rm H}\kern 0.05em$\beta$}} peaks, we construct their cross-dispersion profiles and fit them with a single Gaussian model using the MCMC sampler \texttt{emcee}. For each fit, we extract the full posterior distribution of the Gaussian parameters (amplitude, width, and centroid). Instead of adopting a fixed central credible interval---which can be biased for asymmetric or multi-modal posteriors—--we compute the Highest Posterior Density Interval (HPDI), defined as the narrowest interval containing $68\%$ of the posterior samples. This approach identifies the densest region of probability, providing a more robust measure of uncertainty for skewed or irregular posteriors.

We then measure the width of this tightest $68\%$ interval for the fitted centroid parameter. The HPDI width effectively represents the precision with which the peak location is constrained. To classify a detected enhancement as statistically significant, we require that this width be less than three pixels. 
This threshold ensures that only well-localized, high-confidence peaks are selected as genuine {\hbox{{\rm [O}\kern 0.1em{\sc iii}{\rm ]}}}/{\hbox{{\rm H}\kern 0.05em$\beta$}} enhancements.

By applying this statistical approach to the reduced sample of 87 galaxies, we identify 26 sources that meet our criterion for exhibiting significant localized enhancements in their {\hbox{{\rm [O}\kern 0.1em{\sc iii}{\rm ]}}}/{\hbox{{\rm H}\kern 0.05em$\beta$}} ionization ratios. 
Overall, the detection of such enhancements in approximately $30\%$ of the sample indicates that emission-line ratio profiling can serve as a sensitive probe for identifying compact regions of elevated ionization, potentially tracing obscured or low-luminosity AGN activity. However, since elevated {\hbox{{\rm [O}\kern 0.1em{\sc iii}{\rm ]}}}/{\hbox{{\rm H}\kern 0.05em$\beta$}} ratios can also arise from intense off-nuclear star formation or shock excitation, these spatial peaks are not unique diagnostics of AGN presence. Consequently, while this method efficiently isolates candidates with compact ionization structures, the true AGN fraction within this subset remains uncertain. We subsequently analyze all further results on these 26 sources.

\section{Results:} \label{sec:results}

\subsection{Calculations of Spatial Offsets of {\hbox{{\rm [O}\kern 0.1em{\sc iii}{\rm ]}}}/{\hbox{{\rm H}\kern 0.05em$\beta$}}} \label{sec:techniques-spatialoffset}

To quantify spatial offsets between the peak of the {\hbox{{\rm [O}\kern 0.1em{\sc iii}{\rm ]}}}/{\hbox{{\rm H}\kern 0.05em$\beta$}} flux ratio profile and the peak of the stellar continuum of the galaxy, we compute the difference in spatial pixel positions between the respective peaks for all 26 sources with significant localization. To assess whether these visually identified offsets are statistically significant, we examine the offset posteriors derived for sources exhibiting significant ionization peaks. These posteriors are obtained as part of our procedure for identifying spatially offset emission lines and capture the full uncertainty on the measured displacement between the ionization peak and the continuum peak. We then calculate errorbars ($1\sigma$) for each of the 26 sources. If these errorbars do not include zero, we classify the source as exhibiting a statistically significant spatial offset. Since we use the continuum center as a proxy for the galaxy center and limit our analysis to the central five spatial pixels, the maximum possible spatial offset is $\sim2$ pixels , while the minimum is $0$ pixels (no offset). Applying this criterion, we find that 12 out of the 26 sources ($\sim45\%$) in our sample show credible spatial offsets, with inferred displacements ranging from $\sim0.4$ to $\sim2$ pixels. These pixel-based offsets are converted into physical distances (in kilo-parsecs) using our assumed cosmology. We record the maximum significant measurable spatial offset to be approximately $\sim 1.7$ kpc, while the minimum measurable offset is $\sim 0.24$ kpc. The 12 sources are shown in Table \ref{tab:sigsources}.

We begin by determining the angular diameter distance, $D_A$, for each galaxy using the spectroscopic redshift values from the CEERS catalog. The pixel difference between the ionization peak and the continuum peak is converted to an angular separation using the NIRSpec plate scale of $0.1\arcs$ \citep{nirspec}. We further explore this in Section \ref{localization}.

This offset measurement allows us to evaluate whether the region of highest ionization, traced by elevated {\hbox{{\rm [O}\kern 0.1em{\sc iii}{\rm ]}}}/{\hbox{{\rm H}\kern 0.05em$\beta$}} values, is co-spatial with the galaxy's stellar center or displaced from it. Such significant displacements may indicate the presence of a wandering AGN, potentially resulting from dynamical perturbations such as galaxy mergers or a possible asymmetric starburst. This metric, in conjunction with the ionization strength traced by the flux ratio, also enables us to assess the influence and possible migration of AGNs within their host galaxies during the early stages of galaxy assembly.

To investigate the spatial distribution of of regions of high ionization within our galaxy sample, we compare the locations of the observed {\hbox{{\rm [O}\kern 0.1em{\sc iii}{\rm ]}}}/{\hbox{{\rm H}\kern 0.05em$\beta$}} ionization peaks to the corresponding peak in the galaxy continuum.  

We find that 12 out of the 26 galaxies ($\sim45\%$) in our sample  exhibit significant spatial offsets between the ionization peak and the continuum peak. 
Among these, 6 galaxies show significant offsets of $<1.5$ pixels, while 6 galaxies display larger displacements of $\geq 1.5$ pixels, corresponding to projected physical separations of up to approximately $1.7$ kpc, depending on the individual galaxy redshifts.
The remaining 14 galaxies show no statistically significant offset, as their error bars include zero offset, indicating that the ionization and continuum peaks are consistent with being spatially aligned.

\begin{figure}
    \includegraphics[width=\columnwidth]{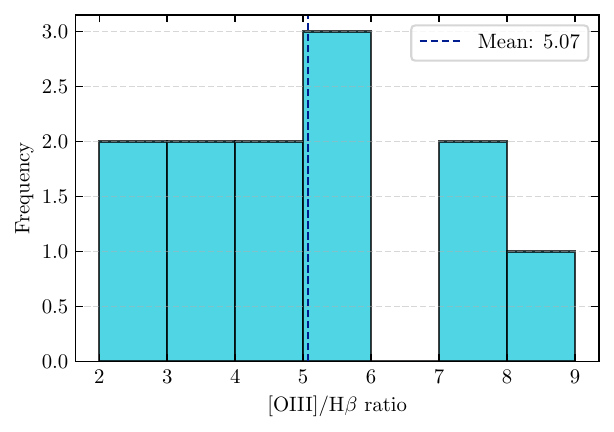}
    \includegraphics[width=\columnwidth]{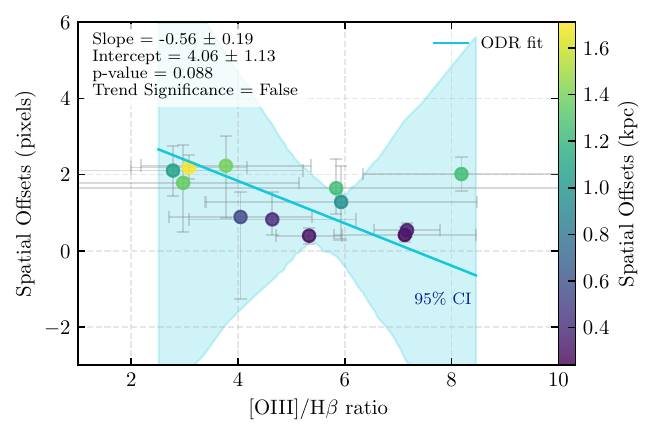}

    \caption{\textbf{Top:} Histogram showing the distribution in the maximum {\hbox{{\rm [O}\kern 0.1em{\sc iii}{\rm ]}}}/{\hbox{{\rm H}\kern 0.05em$\beta$}} flux ratio value for all the sources with significant offsets. This captures the peak of the ionization ratio of each galaxy.
    \textbf{Bottom:} 
    Scatter plot illustrating the correlation between the maximum [OIII]/{\hbox{{\rm H}\kern 0.05em$\beta$}} ionization ratio and the spatial offset in fractional pixels between the [OIII]/{\hbox{{\rm H}\kern 0.05em$\beta$}} peak and the stellar continuum peak. The colorbar shows the corresponding spatial offset in parsecs. We fit the data using an Orthogonal Distance Regression (ODR) model with bootstrap resampling to account for measurement uncertainties in both variables. We also shade region of 95\% bootstrap confidence intervals as shown. The resulting slope is statistically consistent with zero as our p-value is really large, indicating that the trend is not significant. Therefore, we do not find evidence for a correlation between the ionization ratio {\hbox{{\rm [O}\kern 0.1em{\sc iii}{\rm ]}}}/{\hbox{{\rm H}\kern 0.05em$\beta$}} and the spatial offsets.}

\label{fig:spatial_offset_vs_OIII_Hb}
\end{figure}

These results suggest that in a significant fraction of galaxies that exhibit a localized peak of {\hbox{{\rm [O}\kern 0.1em{\sc iii}{\rm ]}}}/{\hbox{{\rm H}\kern 0.05em$\beta$}} ratio, the regions of strongest ionization, are not co-located with the galaxy's continuum based center. The six galaxies exhibiting a $\geq 1.5$-pixel offset represent a particularly interesting subset, as these displacements could imply dynamically significant movement of the central black hole. Such phenomena may result from recent merger events or interactions that displace the black hole from its equilibrium position at the galaxy center \citep{priya_smbh_wander, weller_bh_wandering, agn_wander_jeon, sinking_smbh}.

To evaluate the relationship between these spatial offsets and the ionization strength of {\hbox{{\rm [O}\kern 0.1em{\sc iii}{\rm ]}}}/{\hbox{{\rm H}\kern 0.05em$\beta$}}, we plot the {\hbox{{\rm [O}\kern 0.1em{\sc iii}{\rm ]}}}/{\hbox{{\rm H}\kern 0.05em$\beta$}} ratio versus spatial offset (in sub-pixels) in Figure \ref{fig:spatial_offset_vs_OIII_Hb}. We do not find any significant trends between the extent of the spatial offset (measured between the {\hbox{{\rm [O}\kern 0.1em{\sc iii}{\rm ]}}}/{\hbox{{\rm H}\kern 0.05em$\beta$}} peak and the stellar continuum peak) and the maximum {\hbox{{\rm [O}\kern 0.1em{\sc iii}{\rm ]}}}/{\hbox{{\rm H}\kern 0.05em$\beta$}} ratio for each source. We fit an Orthogonal Distance Regression (ODR) model with bootstrap resampling which yields a slope of $m = -0.56 \pm 0.19$ and an intercept of $c = 4.06 \pm 1.13$, with a two-sided bootstrap p-value of $p = 0.088$. This confirms that the slope is statistically consistent with zero, and therefore no significant correlation is detected at the 95\% confidence level.

\begin{table*}
\centering
\caption{Sources with Significant Offsets}
\label{tab:sigsources}
\scriptsize
\begin{tabular}{ccccccc}
\toprule
MSA\_ID & ID & RA & Dec & Prism\_z & Offset (pixels) & $R_{\rm eff}$ (arcsec) \\
\midrule
3    & CEERS\_4774 & 215.005189  & 52.99658  & 8.005500 & 0.824949264526367 & 0.08646 \\
1065 & EGS\_35576  & 215.1168542 & 53.0010814 & 6.1907783 & 0.416800498962402 & -- \\
1163 & EGS\_39047  & 214.9904678 & 52.9719902 & 7.475 & 2.10724449157715 & -- \\
1620 & EGS\_16658  & 215.087173 & 53.002892 & 5.308 & 0.888683319091797 & -- \\
1699 & EGS\_13957  & 215.0533477 & 52.9648857 & 5.078 & 0.543119430541992 & -- \\
2552 & CEERS\_24867  & 214.8385354 & 52.8850214 & 3.0915 & 2.18902206420898 & 0.15732 \\
4210 & EGS\_7668  & 215.2372077 & 53.0610871 & 5.2607015 & 2.01566791534424 & -- \\
11441 & EGS\_7960  & 215.1779777 & 53.0210701 & 3.39 & 1.78472328186035 & -- \\
11798 & EGS\_11393  & 215.1362241 & 53.0098358 & 3.5293 & 0.396694183349609 & -- \\
12361 & CEERS\_55160  & 214.8864626 & 52.8658322 & 3.28093 & 1.64390897750855 & 0.11001 \\
80072 & CEERS\_87103  & 214.89085 & 52.813941 & 5.2712 & 2.23380565643311 & 0.05439 \\
84534 & CEERS\_35033  & 214.71257 & 52.746281 & 3.464 & 1.28247928619385 & 0.10541999 \\
\bottomrule
\end{tabular}
\end{table*} 

\subsection{Spatial Variations in {\hbox{{\rm [O}\kern 0.1em{\sc iii}{\rm ]}}}/{\hbox{{\rm H}\kern 0.05em$\beta$}} Ratio} \label{localization}

\begin{figure}
    \includegraphics[width=\columnwidth]{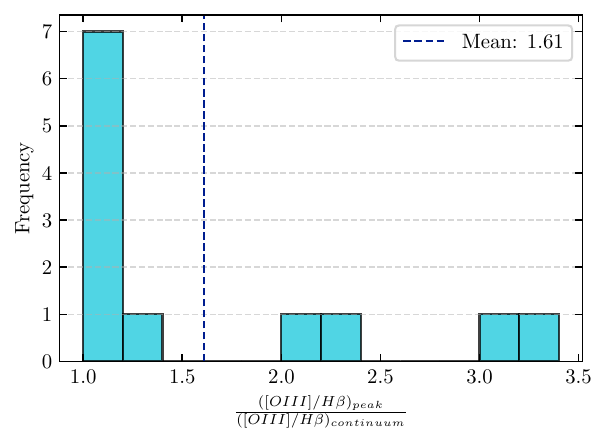}
    
    \includegraphics[width=\columnwidth]{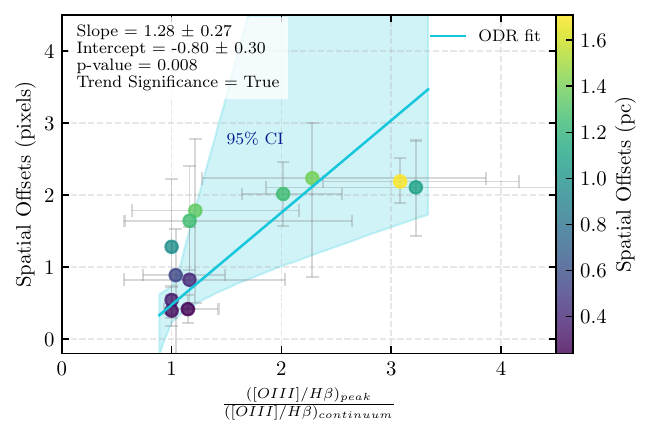}
    
    \caption{\textbf{Top:} Histogram showing the variation in the {\hbox{{\rm [O}\kern 0.1em{\sc iii}{\rm ]}}}/{\hbox{{\rm H}\kern 0.05em$\beta$}} flux ratio, defined as the difference between the maximum {\hbox{{\rm [O}\kern 0.1em{\sc iii}{\rm ]}}}/{\hbox{{\rm H}\kern 0.05em$\beta$}} value and the value at the location of the stellar continuum peak ($\frac{\text{({\hbox{{\rm [O}\kern 0.1em{\sc iii}{\rm ]}}}/H}\beta)_{\text{peak}}}{\text{({\hbox{{\rm [O}\kern 0.1em{\sc iii}{\rm ]}}}/H}\beta)_{\text{continuum}}}$). This metric, computed using Equation~\ref{eq:ion_var_eq}, captures the ionization ratio variation across the central five spatial pixels of each galaxy in the 12 sources with significant spatial offsets. \textbf{Bottom:} Scatter plot showing the relationship between the ionization ratio variation (as shown in the top panel) and the spatial offset in sub pixels ($0$--$2$ pixels). The colorbar indicates the spatial offset in parsecs. We further fit the data using an Orthogonal Distance Regression (ODR) model with bootstrap resampling to account for measurement uncertainties in both variables shown by the blue line fit. We also shade region of 95\% bootstrap confidence intervals as shown. The resulting slope is statistically greater than zero as our p-value is small, indicating that the
    trend is significant; however, we note that systems with larger ionization ratio variations are more easily identified as having a distinct {\hbox{{\rm [O}\kern 0.1em{\sc iii}{\rm ]}}}/{\hbox{{\rm H}\kern 0.05em$\beta$}} peak, which may bias the sample toward higher-contrast sources and enhance the observed correlation.}

\label{fig:ionization_variation}
\end{figure}

To further investigate the relationship between ionization strength and potential AGN localization through spatial offsets, we analyze the variation in the {\hbox{{\rm [O}\kern 0.1em{\sc iii}{\rm ]}}}/{\hbox{{\rm H}\kern 0.05em$\beta$}} flux ratio across the spatial axis of each galaxy. Specifically, we quantify how much the {\hbox{{\rm [O}\kern 0.1em{\sc iii}{\rm ]}}}/{\hbox{{\rm H}\kern 0.05em$\beta$}} ratio peak deviates from the {\hbox{{\rm [O}\kern 0.1em{\sc iii}{\rm ]}}}/{\hbox{{\rm H}\kern 0.05em$\beta$}} level at the galaxy's stellar light-weighted center (continuum peak). This comparison is motivated by the hypothesis that spatially offset ionization peaks may be associated with more intense, localized ionizing sources---potentially AGNs---that are not co-located with the galaxy's morphological center \citep{priya_smbh_wander, weller_bh_wandering, agn_wander_jeon}.

For each galaxy in our sample of sources with significant spatial offsets, we identify the maximum spatial value of the {\hbox{{\rm [O}\kern 0.1em{\sc iii}{\rm ]}}}/{\hbox{{\rm H}\kern 0.05em$\beta$}} flux ratio and compare it to the corresponding ratio value at the continuum peak, which serves as a reference point for the galaxy center. The ratio of these two values provides a measure of the ionization variation, defined as:
\begin{equation}\label{eq:ion_var_eq}
    \text{Ionization Variation} = \frac{\text{({\hbox{{\rm [O}\kern 0.1em{\sc iii}{\rm ]}}}/{\hbox{{\rm H}\kern 0.05em$\beta$}}})_{\text{peak}}}{\text{({\hbox{{\rm [O}\kern 0.1em{\sc iii}{\rm ]}}}/{\hbox{{\rm H}\kern 0.05em$\beta$}}})_{\text{continuum}}}
\end{equation}

This metric captures the relative enhancement of ionization in spatially distinct regions and allows us to investigate whether displaced peaks correspond to relatively stronger ionization activity. We then compare the calculated Ionization Variations with the measured spatial offsets, expressed in both sub-pixels and parsecs, for each galaxy. By doing so, we assess whether a systematic relationship exists between the relative ionization strengths and the physical displacement from the galaxy center.

To evaluate this relationship, we plot the {\hbox{{\rm [O}\kern 0.1em{\sc iii}{\rm ]}}}/{\hbox{{\rm H}\kern 0.05em$\beta$}} variation against the spatial offset. The result, shown in Figure~\ref{fig:ionization_variation}, reveals a moderately positive trend between the magnitude of the spatial offset and the degree of ionization variation, quantified as ($\frac{\text{({\hbox{{\rm [O}\kern 0.1em{\sc iii}{\rm ]}}}/H}\beta)_{\text{peak}}}{\text{({\hbox{{\rm [O}\kern 0.1em{\sc iii}{\rm ]}}}/H}\beta)_{\text{continuum}}}$). We fit an Orthogonal Distance Regression (ODR) model that accounts for uncertainties in both variables, and estimate the uncertainty in the fit parameters using bootstrap resampling. This yields a slope of $m = 1.28 \pm 0.27$ and an intercept of $c = -0.80 \pm 0.30$, with a two-sided bootstrap p-value of $p = 0.008$. These results indicate that the slope is statistically greater than zero, confirming a significant correlation at the 95\% confidence level. Overall, galaxies with larger spatial offsets tend to show stronger enhancements in their peak ionization ratios relative to their continuum values.

This trend suggests that spatially offset sources, those in which the {\hbox{{\rm [O}\kern 0.1em{\sc iii}{\rm ]}}}/{\hbox{{\rm H}\kern 0.05em$\beta$}} peak is displaced from the galaxy continuum peak, are more likely to harbor localized ionizing sources. The positive correlation potentially supports the interpretation that such off-centered AGNs could be more efficient at ionizing the surrounding gas, thereby producing stronger {\hbox{{\rm [O}\kern 0.1em{\sc iii}{\rm ]}}}/{\hbox{{\rm H}\kern 0.05em$\beta$}} enhancements compared to AGNs located at the galaxy center.

Furthermore, this observed variation may reflect asymmetric feedback processes, where displaced AGNs interact differently with the interstellar medium than centrally located ones \citep{greene_offset_AGN_accreting, AGN_offset_GA_NIFS}. Such feedback could lead to modified star formation histories, ultimately affecting the galaxy’s evolutionary journey.

\subsection{OHNO Diagram}
To further assess the ionization mechanisms within our sample, we construct an OHNO diagnostic plot, which compares the 
{\hbox{{\rm [O}\kern 0.1em{\sc iii}{\rm ]}}}/{\hbox{{\rm H}\kern 0.05em$\beta$}} and {\hbox{{\rm [Ne}\kern 0.1em{\sc iii}{\rm ]}}}/{\hbox{{\rm [O}\kern 0.1em{\sc ii}{\rm ]}}} flux ratios \citep{Backhaus_OHNO}. This diagram was originally designed to distinguish star-forming galaxies from X-ray AGN and [Ne V] emitters at $z\sim1$, with the empirical boundary calibrated in \citet{Backhaus_OHNO}. The utility of this diagnostic arises from the fact that, at low redshifts, star-forming regions typically have lower ionization parameters and softer ionizing spectra, such that they separate from AGN along the diagonal trend of the diagram. 

Both {\hbox{{\rm [O}\kern 0.1em{\sc iii}{\rm ]}}}/{\hbox{{\rm H}\kern 0.05em$\beta$}} and {\hbox{{\rm [Ne}\kern 0.1em{\sc iii}{\rm ]}}}/{\hbox{{\rm [O}\kern 0.1em{\sc ii}{\rm ]}}} are primarily sensitive to the ionization parameter at comparable ionization energies. As a result, galaxies with elevated ionization conditions tend to move coherently along both axes of the OHNO diagram. In this sense, the correlated axes are advantageous when identifying highly ionized systems, as extreme sources are reinforced in both diagnostics. However, real galaxies do not lie on a perfectly diagonal relation; there exists measurable scatter in {\hbox{{\rm [Ne}\kern 0.1em{\sc iii}{\rm ]}}}/{\hbox{{\rm [O}\kern 0.1em{\sc ii}{\rm ]}}} ratios that is not yet fully understood and may reflect variations in metallicity, ionizing spectra, or nebular conditions.

At higher redshifts, the primary caveat arises not from the axis correlation itself, but from the evolving ionization conditions of galaxies. Recent work has shown that the empirical separation between star formation and AGN becomes less robust at $z \gtrsim 3$ \citep[e.g.,][]{Backhaus_OHNO, bren_ratios_1, bren_ratios_2, nikko_ionization, nikko_OHNO}, likely due to harder stellar radiation fields, lower metallicities, and elevated ionization parameters in early galaxies. Classifications based on low-$z$ boundaries should therefore be treated with caution. Nevertheless, we employ the OHNO diagram as a comparative framework, as it remains one of the most widely used empirical diagnostics for distinguishing ionization sources and provides a valuable baseline for assessing how high-redshift systems deviate from well-characterized local trends. We again choose not to perform dust corrections for these ratios as the corresponding emission lines for each of the ratios are closely separated in wavelength such that the relative corrections are negligible.

Before constructing the OHNO plot, we apply a second signal-to-noise ratio cut to the narrow emission lines {\hbox{{\rm [Ne}\kern 0.1em{\sc iii}{\rm ]}}} and {\hbox{{\rm [O}\kern 0.1em{\sc ii}{\rm ]}}}, requiring $\rm SNR\geq2$. This criterion is applied uniformly across all 26 sources that have significant O~III/{\hbox{{\rm H}\kern 0.05em$\beta$}} peaks and doing so reduces the sample to 16 sources with robust OHNO values. Out of these 16 sources, 9 sources have significant spatial offsets.

\begin{figure}
    \includegraphics[width=\columnwidth]{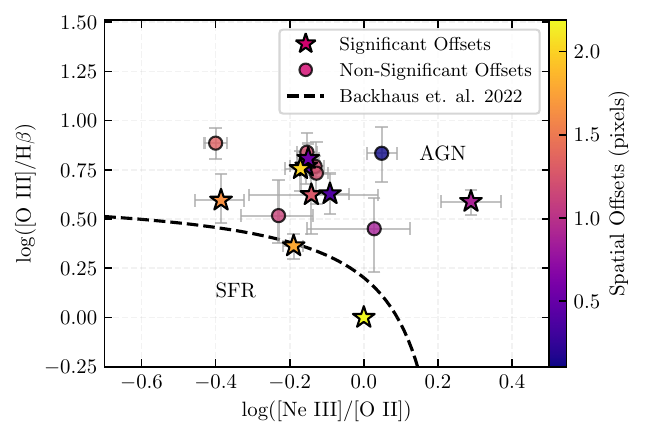}
    \includegraphics[width=\columnwidth]{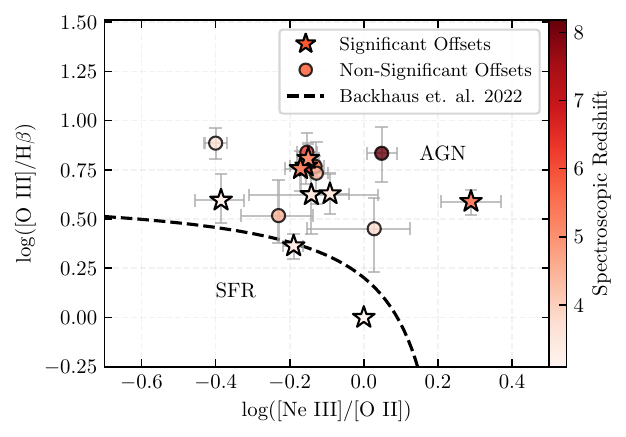}
    \caption{OHNO Diagnostic Plot which shows emission lines ratios of log({\hbox{{\rm [O}\kern 0.1em{\sc iii}{\rm ]}}}/{\hbox{{\rm H}\kern 0.05em$\beta$}}) and log({\hbox{{\rm [Ne}\kern 0.1em{\sc iii}{\rm ]}}}/{\hbox{{\rm [O}\kern 0.1em{\sc ii}{\rm ]}}}). The dashed line represents the OHNO diagnostic boundary from \citet{Backhaus_OHNO}, which separates AGN-dominated sources from those dominated by star formation regions (SFR).  
    \textbf{Top:} The colorbar encodes the spatial offset in pixel units between the peaks of the ionization ratio and the stellar continuum. 
    \textbf{Bottom:} The colorbar encodes the redshift of the sources. Star-shaped points denote sources with statistically significant offsets, while circle-shaped points represent sources without statistically significant offsets.
    We find that most sources lie in the AGN-dominated region of the diagram and a subset of sources have high {\hbox{{\rm [Ne}\kern 0.1em{\sc iii}{\rm ]}}}/{\hbox{{\rm [O}\kern 0.1em{\sc ii}{\rm ]}}} ratio values which could be suggestive of narrow line AGN activity with the caveat that at $z \gtrsim 2$ the calibration of this diagnostic is uncertain and may be offset by harder ionizing conditions.
}

\label{fig:OHNO}
\end{figure}

In our sample, most galaxies occupy the AGN-dominated region of the OHNO diagram as shown in Figure \ref{fig:OHNO}, which is consistent with galaxies at higher redshift since the OHNO diagram's distinguishing power is diminished at high redshift \citep{nikko_OHNO}. Furthermore, some sources that have a higher {\hbox{{\rm [Ne}\kern 0.1em{\sc iii}{\rm ]}}}/{\hbox{{\rm [O}\kern 0.1em{\sc ii}{\rm ]}}} value, located towards the very right of the clump of sources, could suggest that these sources have traces of narrow-line AGN activity. This distribution reinforces the interpretation that displaced ionization regions, potentially tracing AGN activity, play a significant role in the ionization properties of the selected high-redshift galaxies. However, it also raises questions about the robustness of the OHNO diagnostic at $z > 3$, where elevated {\hbox{{\rm [O}\kern 0.1em{\sc iii}{\rm ]}}}/{\hbox{{\rm H}\kern 0.05em$\beta$}} ratios can sometimes result from extreme star-forming conditions rather than purely AGN activity. As discussed in \citet{Backhaus_OHNO, bren_ratios_2, nikko_OHNO}, such high-redshift effects must be carefully considered when applying traditional diagnostics. 

In the bottom panel of Figure \ref{fig:OHNO}, we present the OHNO diagram with the colorbar indicating source redshift. We find that higher-redshift galaxies tend to lie above the empirical AGN/SF demarcation line, which could either signal an increased prevalence of possible AGN activity or reflect an evolution in the typical {\hbox{{\rm [O}\kern 0.1em{\sc iii}{\rm ]}}}/{\hbox{{\rm H}\kern 0.05em$\beta$}} ratios at early epochs. The latter explanation is consistent with elevated ionization parameters in star-forming regions at high redshift, as has also been discussed for the Baldwin, Philips, \& Terlevich (BPT) diagram in the context of photoionization modeling \citep[e.g.,][]{bpt81, Kewley_BPT_2013}. Because the AGN/SF separation line was defined at $z\sim1.7$, its direct applicability to higher redshifts is uncertain as star-forming galaxies are expected to migrate toward the AGN region as their ionizing spectra harden, and narrow-line AGN likely contaminate the star-forming locus, complicating empirical recalibrations \citep{Backhaus_OHNO,bren_ratios_2}. These caveats highlight the limitations of using OHNO for binary classifications at high redshift. Nevertheless, the diagram remains useful for relative comparisons of gas ionization properties within a high-$z$ sample, even if individual classifications should be treated with caution.

Assessing the values of the OHNO diagnostic plot and its robustness could show some possible connections between off-center AGN activity and enhanced ionization conditions. This offers an avenue for future investigation into the relationship between AGN dynamics and ionization strength at early cosmic times.

\section{Conclusions} \label{sec:summary}

In this study, we investigate the spatially resolved ionization properties of 87 high-redshift ($3 < z < 8$) galaxies with high S/N observed with JWST/NIRSpec PRISM as part of the CEERS survey \citep{CEERS2025}. By focusing on the {\hbox{{\rm [O}\kern 0.1em{\sc iii}{\rm ]}}}/{\hbox{{\rm H}\kern 0.05em$\beta$}} flux ratio, a well-established diagnostic of ionizing activity, we identify significant ionization peaks and assess their spatial alignment with the stellar continuum centers of the galaxies. Our primary goal is to determine whether elevated ionization and spatial displacements are indicative of AGN activity and to explore black hole–galaxy co-evolution in the early universe.

Our main results can be summarized as follows:
\begin{enumerate}
    \item \textbf{Significant {\hbox{{\rm [O}\kern 0.1em{\sc iii}{\rm ]}}}/{\hbox{{\rm H}\kern 0.05em$\beta$}} Flux Ratio Peaks:} $\sim30\%$ of the sample (26 out of 87 galaxies) show localized enhancement in peaks in their {\hbox{{\rm [O}\kern 0.1em{\sc iii}{\rm ]}}}/{\hbox{{\rm H}\kern 0.05em$\beta$}} profiles, consistent with compact ionizing sources which could have AGN activity. These profiles exhibit positive values for the significant peak statistic and have central enhancements that taper toward the outskirts, suggesting a dominant, spatially confined ionization mechanism.

    \item \textbf{Spatial Offsets:} 12 out of the 26 galaxies ($\sim46\%$) display measurable spatial offsets between the {\hbox{{\rm [O}\kern 0.1em{\sc iii}{\rm ]}}}/{\hbox{{\rm H}\kern 0.05em$\beta$}} ionization peak and the continuum (light-weighted stellar) peak. 
    These displacements may be indicative of off-nuclear or wandering AGNs or isolated off-center star formation. 
    
    \item \textbf{Trends in Spatial Offsets and {\hbox{{\rm [O}\kern 0.1em{\sc iii}{\rm ]}}}/{\hbox{{\rm H}\kern 0.05em$\beta$}} ratios}: We observe no direct trends between spatial ionization offsets and {\hbox{{\rm [O}\kern 0.1em{\sc iii}{\rm ]}}}/{\hbox{{\rm H}\kern 0.05em$\beta$}}. However, we observe a positive trend between spatial offset sources and variations in {\hbox{{\rm [O}\kern 0.1em{\sc iii}{\rm ]}}}/{\hbox{{\rm H}\kern 0.05em$\beta$}} ($\frac{\text{({\hbox{{\rm [O}\kern 0.1em{\sc iii}{\rm ]}}}/{\hbox{{\rm H}\kern 0.05em$\beta$}}})_{\text{peak}}}{\text{({\hbox{{\rm [O}\kern 0.1em{\sc iii}{\rm ]}}}/{\hbox{{\rm H}\kern 0.05em$\beta$}}})_{\text{continuum}}}$). This suggests that galaxies with larger spatial displacements tend to exhibit greater relative ionization enhancement, suggesting that more intense AGN-driven ionization may be occurring away from the stellar center.

    \item \textbf{OHNO Diagnostic Analysis:} Most galaxies in our sample occupy the AGN-designated region of the OHNO diagram, consistent with their elevated {\hbox{{\rm [O}\kern 0.1em{\sc iii}{\rm ]}}}/{\hbox{{\rm H}\kern 0.05em$\beta$}} ratios and showing signs of redshift evolution. Some sources as seen in the plot with extreme {\hbox{{\rm [Ne}\kern 0.1em{\sc iii}{\rm ]}}}/{\hbox{{\rm [O}\kern 0.1em{\sc ii}{\rm ]}}} values may indicate traces of narrow-line AGN activity. However, as the OHNO boundary was calibrated at $z\sim1.7$ and the diagnostic becomes unreliable at $z>3$, elevated ionization from star-forming regions can mimic AGN-like values. Thus, while useful for relative comparisons, the OHNO diagram should be applied with caution at high redshift \citep{Backhaus_OHNO, bren_ratios_2, nikko_OHNO}.

\end{enumerate}

These findings suggest that elevated {\hbox{{\rm [O}\kern 0.1em{\sc iii}{\rm ]}}}/{\hbox{{\rm H}\kern 0.05em$\beta$}} ratios—particularly when accompanied by spatial offsets—may be associated with AGN activity, although they do not provide direct or unique confirmation of this scenario. The inferred projected offsets of $\sim0.24$--$1.7$ kpc are also broadly consistent with the range of spatial displacements predicted in recent theoretical studies of wandering black holes. Simulations find offset populations spanning a wide range of scales, from merger-driven displacements of order tens to hundreds of parsecs within galaxy centers \citep{chen2022_imbh} to kpc- and even tens-of-kpc-scale wanderers associated with hierarchical assembly, satellite accretion, and long-lived dynamical perturbations \citep{ricarte2021_imbh_mergers,DiMatteo_2023_imbh}. Our measurements therefore occupy an intermediate regime that is compatible with theoretical expectations for wandering SMBHs, while remaining substantially smaller than the most extreme offsets predicted in some cosmological simulations. Given that our measurements represent projected separations and are limited to a single spatial dimension, the intrinsic three-dimensional offsets could be larger \citep{ricarte2021_imbh_mergers,priya_smbh_wander, weller_bh_wandering, agn_wander_jeon}.

At the same time, we emphasize that spatially localized ionization peaks alone cannot unambiguously distinguish between AGN-driven and stellar-driven ionization. Compact, off-center star-forming regions—particularly in low-metallicity environments with hard radiation fields—can produce elevated {\hbox{{\rm [O}\kern 0.1em{\sc iii}{\rm ]}}}/{\hbox{{\rm H}\kern 0.05em$\beta$}} ratios that mimic AGN-like signatures, while shocks associated with feedback or dynamical disturbances may also contribute. The recovery of known broad-line AGNs using our method and the prevalence of our sources in the AGN-designated region of the OHNO diagram lend support to an AGN interpretation for at least a subset of the sample, but alternative explanations remain viable on a case-by-case basis.

We find no significant correlation between the amplitude of the spatial offset and the peak {\hbox{{\rm [O}\kern 0.1em{\sc iii}{\rm ]}}}/{\hbox{{\rm H}\kern 0.05em$\beta$}} ratio, consistent with expectations that ionization strength depends primarily on accretion rate, gas density, and geometry rather than on the black hole’s precise location within the galaxy, although the connection is likely more complex since spatially offset AGNs would still require sufficiently clumpy gas away from the galaxy center to sustain accretion. However, the positive correlation between spatial offset and the relative ionization enhancement with respect to the stellar continuum center suggests that displaced ionizing sources more strongly dominate their local environments. This behavior is naturally explained in scenarios where off-nuclear AGNs produce compact, localized ionized regions that stand out against a lower-ionization stellar background.

If a fraction of the spatially offset ionization peaks identified in this work are indeed powered by wandering AGNs, this has important implications for early SMBH–galaxy co-evolution. Off-center accretion may deposit energy and momentum into the interstellar medium in a more asymmetric manner than centrally located AGNs, potentially influencing gas dynamics, star formation efficiency, disk stability, and the coupling of feedback to the host galaxy \citep{trump}. The apparent prevalence of such systems further suggests that SMBH growth and stellar mass assembly may not always be tightly coupled spatially during the early stages of galaxy evolution.

Overall, this work provides new insight into the spatial complexity of ionizing sources in high-redshift galaxies. Our spatially resolved emission-line analysis reveals ionization enhancements that may occur at significant distances from the stellar centers of their host galaxies, consistent with—but not uniquely indicative of—off-nuclear AGN activity. Future observations, particularly deeper data and two-dimensional mapping with the NIRSpec IFU, will be crucial for disentangling AGN-driven ionization from stellar and shock-driven processes and for fully characterizing the role of spatially displaced accretion in shaping early galaxy evolution.

\section*{Acknowledgments:} 

U.T. thanks the University of Texas at Austin Galaxy Evolution Vertically Integrated Projects (GEVIP) undergraduate research program for providing sustained mentorship, collaborative research training, and institutional support that were instrumental in the conception, development, and successful execution of this project. A.T. acknowledges support from the UT Austin College of Natural Sciences.

This work is based on observations made with the NASA/ESA/CSA \textit{James Webb Space Telescope}, obtained at the Space Telescope Science Institute, which is operated by the Association of Universities for Research in Astronomy, Incorporated, under NASA contract NAS5-03127. The data were obtained from the Mikulski Archive for Space Telescopes (MAST) at the Space Telescope Science Institute. 
These observations are associated with program \#1345, and can be accessed via \citet{data_ceers_obs}.


\newpage

\bibliographystyle{aasjournalv7.bst}
\bibliography{biblibrary.bib}

\end{document}